\documentclass[aps,prl,reprint,nobibnotes,superscriptaddress]{revtex4-2}
\usepackage{graphicx}
\graphicspath{ {} }
\usepackage[load-configurations=abbreviations,uncertainty-mode = separate]{siunitx}
\usepackage[version=4]{mhchem}
\usepackage[dvipsnames]{xcolor}
\usepackage{amssymb}
\usepackage{footmisc}
\definecolor{orange}{rgb}{1,0.5,0}

\usepackage{verbatim}

\begin{document}
\title{Nanoscale transient polarization gratings}

\author{Laura Foglia}
\email[]{laura.foglia@elettra.eu}
\affiliation{Elettra - Sincrotrone Trieste S.C.p.A., S.S. 14 km 163.5 in Area Science Park, 34192 Trieste, Italy}
\author{Bj\"orn Wehinger}
\altaffiliation[Now at: ]{European Synchrotron Radiation Facility, 71 Avenue des Martyrs, Grenoble 38000, France}
\affiliation{Elettra - Sincrotrone Trieste S.C.p.A., S.S. 14 km 163.5 in Area Science Park, 34192 Trieste, Italy}
\affiliation{Department of Molecular Sciences and Nanosystems, Ca’ Foscari University of Venice, 30172 Venezia, Italy}
\author{Giovanni Perosa}
\affiliation{Elettra - Sincrotrone Trieste S.C.p.A., S.S. 14 km 163.5 in Area Science Park, 34192 Trieste, Italy}
\affiliation{Department of Physics, Universit\'{a} degli Studi di Trieste, 34127 Trieste, Italy}
\author{Riccardo Mincigrucci}
\affiliation{Elettra - Sincrotrone Trieste S.C.p.A., S.S. 14 km 163.5 in Area Science Park, 34192 Trieste, Italy}
\author{Enrico Allaria}
\affiliation{Elettra - Sincrotrone Trieste S.C.p.A., S.S. 14 km 163.5 in Area Science Park, 34192 Trieste, Italy}
\author{Francesco Armillotta}
\affiliation{Department of Physics, Universit\'{a} degli Studi di Trieste, 34127 Trieste, Italy}
\affiliation{Institut de Physique des Nanostructures, Ecole Polytechnique F\'{e}d\'{e}rale de Lausanne, CH-1015 Lausanne, Switzerland.}
\author{Alexander Brynes}
\affiliation{Elettra - Sincrotrone Trieste S.C.p.A., S.S. 14 km 163.5 in Area Science Park, 34192 Trieste, Italy}
\author{Riccardo Cucini}
\affiliation{Istituto Officina dei Materiali, Consiglio Nazionale delle Ricerche, Area Science Park, 34192 Trieste, Italy}
\author{Dario De Angelis}
\affiliation{Elettra - Sincrotrone Trieste S.C.p.A., S.S. 14 km 163.5 in Area Science Park, 34192 Trieste, Italy}
\author{Giovanni De Ninno}
\affiliation{Elettra - Sincrotrone Trieste S.C.p.A., S.S. 14 km 163.5 in Area Science Park, 34192 Trieste, Italy}
\affiliation{Laboratory of Quantum Optics, University of Nova Gorica, 5270 Ajdovščina, Slovenia}
\author{W. Dieter Engel}
\affiliation{Max-Born-Institute for Nonlinear Optics and Short Pulse Spectroscopy, 12489 Berlin, Germany}
\author{Danny Fainozzi}
\author{Luca Giannessi}
\affiliation{Elettra - Sincrotrone Trieste S.C.p.A., S.S. 14 km 163.5 in Area Science Park, 34192 Trieste, Italy}
\author{Nupur N. Khatu}
\affiliation{Elettra - Sincrotrone Trieste S.C.p.A., S.S. 14 km 163.5 in Area Science Park, 34192 Trieste, Italy}
\affiliation{Department of Molecular Sciences and Nanosystems, Ca’ Foscari University of Venice, 30172 Venezia, Italy}
\affiliation{European XFEL, Holzkoppel 4, 22869 Schenefeld, Germany}
\author{Simone Laterza}
\affiliation{Elettra - Sincrotrone Trieste S.C.p.A., S.S. 14 km 163.5 in Area Science Park, 34192 Trieste, Italy}
\affiliation{Department of Physics, Universit\'{a} degli Studi di Trieste, 34127 Trieste, Italy}
\author{Ettore Paltanin}
\affiliation{Elettra - Sincrotrone Trieste S.C.p.A., S.S. 14 km 163.5 in Area Science Park, 34192 Trieste, Italy}
\affiliation{Department of Physics, Universit\'{a} degli Studi di Trieste, 34127 Trieste, Italy}
\author{Jacopo Stefano Pelli-Cresi}
\author{Giuseppe Penco}
\affiliation{Elettra - Sincrotrone Trieste S.C.p.A., S.S. 14 km 163.5 in Area Science Park, 34192 Trieste, Italy}
\author{Denny Puntel}
\affiliation{Department of Physics, Universit\'{a} degli Studi di Trieste, 34127 Trieste, Italy}
\author{Primo\v{z} Rebernik Ribi\v{c}}
\affiliation{Elettra - Sincrotrone Trieste S.C.p.A., S.S. 14 km 163.5 in Area Science Park, 34192 Trieste, Italy}
\author{Filippo Sottocorona}
\affiliation{Elettra - Sincrotrone Trieste S.C.p.A., S.S. 14 km 163.5 in Area Science Park, 34192 Trieste, Italy}
\affiliation{Department of Physics, Universit\'{a} degli Studi di Trieste, 34127 Trieste, Italy}
\author{Mauro Trov\`{o}}
\affiliation{Elettra - Sincrotrone Trieste S.C.p.A., S.S. 14 km 163.5 in Area Science Park, 34192 Trieste, Italy}
\author{Clemens von Korff Schmising}
\author{Kelvin Yao}
\affiliation{Max-Born-Institute for Nonlinear Optics and Short Pulse Spectroscopy, 12489 Berlin, Germany}
\author{Claudio Masciovecchio}
\affiliation{Elettra - Sincrotrone Trieste S.C.p.A., S.S. 14 km 163.5 in Area Science Park, 34192 Trieste, Italy}
\author{Stefano Bonetti}
\affiliation{Department of Molecular Sciences and Nanosystems, Ca’ Foscari University of Venice, 30172 Venezia, Italy}
\author{Filippo Bencivenga}
\affiliation{Elettra - Sincrotrone Trieste S.C.p.A., S.S. 14 km 163.5 in Area Science Park, 34192 Trieste, Italy}


\date{\today}

\begin{abstract}
We present the generation of transient polarization gratings at the nanoscale, achieved using a tailored accelerator configuration of the FERMI free electron laser. We demonstrate the capabilities of such a transient polarization grating by comparing its induced dynamics with the ones triggered by a more conventional intensity grating on a thin film ferrimagnetic alloy. While the signal of the intensity grating is dominated by the thermoelastic response of the system, such a contribution is suppressed in the case of the polarization grating. This exposes helicity-dependent magnetization dynamics that have so-far remained hidden under the large thermally driven response. We anticipate nanoscale transient polarization gratings to become useful for the study of any physical, chemical and biological systems possessing chiral symmetry.

\end{abstract}


\maketitle
Manipulating light at the nanoscale is a major challenge for modern science, with the potential to unveil fundamental aspects of light-matter interactions and to enable advances in key technologies such as light harvesting, imaging, biosensing or catalysis. In the visible range, nanoscale control of properties of the radiation such as intensity and phase is often achieved via artificial structures with dimensions comparable to or shorter than the wavelength, such as metasurfaces, photonic crystals and plasmonic nanostructures  \cite{Rot2014,Man2019,Bur2009}. Instead, nanoscale control of light polarization remains particularly challenging. Spatially variable inhomogeneous vector beams can for instance be generated by using specially designed metasurfaces \cite{Zha2018}, but a straightforward approach at short wavelengths is yet to be found.

The use of coherent extreme ultraviolet (EUV) and X-ray radiation in a transient grating (TG) scheme provides an alternative way to control light fields at the nanoscale. In fact, the brightness of EUV free electron laser (FEL) pulses allows to efficiently generate sinusoidal patterns of light intensity, without requiring any physical modification of the sample. With spatial periodicity in the \qtyrange[range-units = single]{10}{100}{nm} range \cite{Bencivenga2015a,Ben2019}, such an EUV TG excitation is then capable of driving ultrafast nanoscale dynamics in different kinds of materials in a controlled way. It has recently proven to be an effective tool for investigating the thermal and mechanical properties of matter in a length-scale range previously inaccessible \cite{Foglia2023,Bencivenga2023}. Additionally, the access to core resonances provides unprecedented insights in the electronic and magnetic dynamics at such a lengthscale \cite{Ksenzov2021,Yao2022}.

In this Letter, we show how the EUV TG approach can be used to generate ultrafast modulations of light polarization in the tens of nm scale. This unique capability can be broadly applied to study helicity-dependent dynamics at the nanoscale and with ultrafast time resolution. We show the potential of the polarization transient grating by investigating the ultrafast demagnetization of a thin film ferrimagnetic alloy, where we are able to trigger non-thermal magnetization dynamics that are otherwise hidden by the thermoelastic response.

\begin{figure}[]
\includegraphics[]{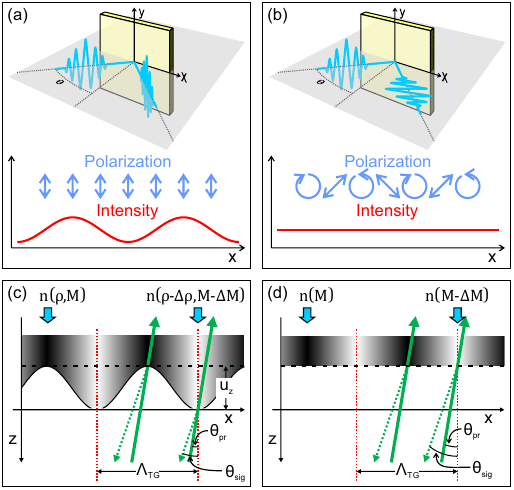}%
\caption{\label{Fig1} (a) PP beams generate an intensity grating with period $\Lambda_{TG}$ and constant polarization. (b) OP beams generate a modulated polarization with the same periodicity and constant intensity. (c) Backward diffraction from an intensity TG resulting from surface modulations $u_z$ and refractive index variations due to changes in both density and magnetization. (d) Backward diffraction from a polarization TG, where the absence of thermal modulation removes both the $u_z$ and $\Delta \rho$ modulations. The signal originates only from refractive index variations associated to the chiral response.}
\end{figure}

In a TG experiment, two pulses of the same wavelength $\lambda_{ex}$ and intensity $I_0$, overlapped in time and space and at a given crossing angle $2\theta$ at the sample, generate a transient interference pattern. If the two excitation beams are parallelly-polarized (PP), as in Figure \ref{Fig1}(a), the intensity at the sample is fully modulated and given by $I_{ex}=2I_0(1+\cos{qx})$, where $q$ is the grating wavevector. The spatial periodicity is given by $\Lambda_{TG}=\lambda_{ex}/2\sin(\theta)=2\pi/q$, while the polarization remains uniform.
Instead, when the two beams are orthogonally-polarized (OP), the polarization is modulated with the same periodicity $\Lambda_{TG}$, ranging from circular left to circular right, while the intensity remains uniform \cite{Eichler1986,Terazima1995,Yang2012,Weber2005}, as depicted in Figure \ref{Fig1}(b). Therefore, analogously to EUV intensity TGs, EUV polarization TGs enable the generation of ultrafast modulations of light polarization with values of $\Lambda_{TG}$ comparable to those of $\lambda_{ex}\approx\qtyrange[range-units=single, range-phrase=-]{10}{100}{nm}$.

Both intensity and polarization gratings can be revealed by the diffraction of an EUV probe beam. The diffraction intensity as a function of the time delay ($\Delta t$) with respect to the EUV TG excitation encodes information on the sample dynamics.
In the case of an EUV intensity TG, after $\approx \qty{100}{fs}$ the electronic grating generated by photo-absorption transfers its energy to the lattice via electron-phonon coupling, leading to a temperature grating \cite{Foglia2023}. This induces a time- and space-dependent density modulation $\Delta\rho(x, \Delta t)$ via thermal expansion and, in magnetic samples, also a magnetization modulation $\Delta M(x, \Delta t)$ \cite{Ksenzov2021}.
These two thermal effects contribute to an effective periodic variation of the refractive index $\Delta n(\rho, M) = n(\rho, M) - n(\rho - \Delta \rho, M - \Delta M)$ between the unexcited areas, which retain equilibrium density $\rho$ and magnetization $M$, and the expanded and demagnetized photo-excited stripes. This periodic variation leads to the diffraction of the EUV probe beam, provided that $\Delta n(\rho, M)\neq 0$ at the wavelength of the probe $\lambda_{pr}$.
Additionally, an intensity grating is associated to a surface displacement, $u_z(x,\Delta t) = 2u_z(\Delta t)(1+\cos{qx})$ (see \ref{Fig1}(c)), resulting from thermal expansion.\footnote{The figure neglects the ultrafast, sub-ps, electronic contribution to $\Delta n$ since the whole discussion concentrates on the \unit{ps} dynamics.}. In forward diffraction, this latter contribution is typically negligible with respect to the signal arising from $\Delta n(\rho, M)$, which can be enhanced via propagation through the bulk of the sample \cite{Foglia2023, Bencivenga2023}. In backward diffraction, on the contrary, the contribution of surface displacement typically dominates and vanishes only in the absence of a temperature grating. 
This is indeed the case of the polarization grating, where the uniform excitation intensity, and thus uniform temperature, prevents both surface displacement and bulk density modulations. Instead, a pure polarization grating can still modulate $\Delta n$ (see \ref{Fig1}(d)), provided that the sample shows a non-thermal chiral response. Such a response can be caused, for instance, by the inverse Faraday effect (IFE) or the magnetization induced by light absorption (MILA) mechanism in magnetic samples \cite{Scheid2022}.
Thus, in backward diffraction and in a magnetic sample, a polarization grating suppresses the $u_z$ contribution and reveals the remaining helicity-dependent magnetic contribution to $\Delta n$. More generally, pure nanoscale polarization gratings can be used to isolate the chiral contributions to $\Delta n$.

We demonstrate this experimentally in a thin film CoGd alloy, where Ksenzov et al. \cite{Ksenzov2021} have previously observed a strong TG signal in forward diffraction when probing resonantly at the Co M-edge. For $\Delta t$ larger than 100s of \unit{fs}, the diffraction intensity is mainly determined by $\Delta M$, i.e. $\Delta n(\rho, M)\approx \frac{\partial n}{\partial M}\Delta M(x,\Delta t) =\Delta n_M(x, \Delta t) \propto 2\Delta M(\Delta t)(1+\cos{qx})/M$ \cite{Ksenzov2021}. That magnetic response has a clear non-sinusoidal time-dependence and should be observed also in backward diffraction. However, the $u_z(x,\Delta t)$ contribution is expected to be dominant, according to previous results \cite{Maznev2021} and considering the typical thermal expansion coefficients of metallic films. The exact temporal evolution of the surface-induced signal is generally unknown a priori. Nevertheless, in these \unit{nm}-thick layered samples on a bulk substrate, it typically consists of a combination of surface acoustic waves (SAW), Lamb modes and leaky waves that are all associated to sinusoidal waveforms in time. Thus, the sinusoidal thermoelastic and the non-sinusoidal magnetic dynamics can be easily discerned on the \unit{ps} timescale. When exciting with OP beams, the suppression of the oscillatory signal and the isolation of the non-sinusoidal signal from $\Delta n_M(x, \Delta t)$ indicate the generation of the nanoscale polarization grating.

Practically however, realizing EUV polarization gratings is not a trivial endeavor. Polarization control after the EUV source has been demonstrated using phase retardation upon reflection off metallic mirrors \cite{vonKorff2017}. Such a setup would allow for independent OP beams, but it sacrifices the overall efficiency and complicates greatly the experimental scheme. On the other hand, polarization control of the EUV source \cite{Allaria2014,Roussel2017,Lutman2016,Deng2014,Schmidt2018,Schneidmiller2018,Li2017,Yakopov2022,Perosa2023} does not allow the simultaneous generation of multiple EUV pulses with different polarization \textit{and direction}. Instead, in this Letter we present a special configuration of the FERMI FEL (Trieste, Italy), which permits to simultaneously obtain a pair of EUV pulses with independent linear polarization emitted along slightly different trajectories. Moreover, it allows to switch from PP to OP beams without changing any other experimental parameter. Further details on the accelerator setup are provided in the Appendix. The FEL is tuned at an operating wavelength of \qty{20.8}{\nm}, to match the Co M-edge resonance. The two FEL beams are sent each on one of the two pump branch lines of the focusing system of the TIMER beamline, which is used to cross them at the sample with an angle $2\theta= \ang{27.6}$, resulting in $\Lambda_{TG} = \qty{43.6}{\nm}$. The TIMER setup is also used to generate a variably-delayed probe beam \cite{Mincigrucci2018} with wavelength $\lambda_{pr}=\qty{20.8}{\nm}$ and linear polarization parallel to the \textit{y}-axis as defined in Fig. \ref{Fig1}.

The sample is a \qty{15}{\nm} thick film of \ce{Co_{0.78}Gd_{0.22}} alloy with perpendicular magnetic anisotropy, deposited within a metallic stack (\qty{1.5}{\nm} \ce{Ta}/\qty{1.5}{\nm} \ce{Pt}/ \qty{15}{\nm} \ce{Co_{0.78}Gd_{0.22}}/ \qty{3}{\nm} \ce{Pt}) on a lead-germanate glass substrate, analogous to the one used in reference \cite{Ksenzov2021} (see Ref. \cite{[{See Supplemental Material at }][{for details on the sample preparation and characterization, and a discussion on the fitting procedure and thermoelastic properties of the sample, which includes Refs. [30,31] }]Suppl} and Ref. \cite{Ceballos2021,Shirakawa1985} therein). The film is magnetized to saturation with the magnetic field normal to the film surface. The excitation fluence at the sample is about \qty{5}{\milli J\per\cm\squared}, the penetration depth at \qty{20.8}{nm} in the magnetic layer is calculated to be \qty{16}{nm} and we estimate a local sample heating due to FEL excitation of about \qty{400}{\K}. 
Each acquisition is integrated over 1500 shots per delay point with a FEL repetition rate of \qty{50}{Hz} and a pulse duration of $\approx \qty{40}{\fs}$. We measured the total diffracted signal without polarization analysis. 

\begin{figure}
\includegraphics{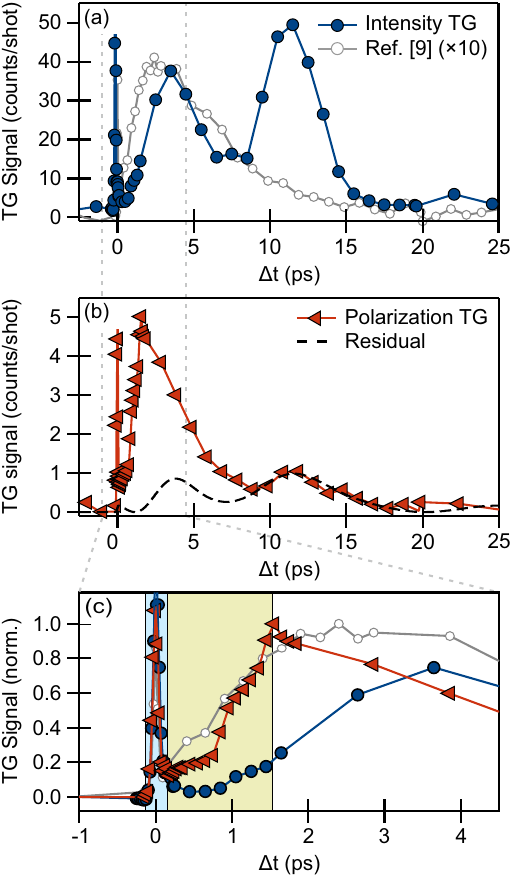}%
\caption{\label{Fig2} (a) Intensity grating. Blue circles: backward-diffracted TG signal from PP beams. Grey circles: Ref. \cite{Ksenzov2021}. (b) Polarization gratings. Red triangles: backward-diffracted signal from OP beams. Dashed line: residual sinusoidal component (see text for discussion). (c) Zoom-in of the early-time dynamics, evidencing the difference in the functional dependence of the signals, normalized to their maximum. }
\end{figure}

Figure \ref{Fig2}(a) depicts the backward-diffracted intensity TG signal (PP pump beams) as blue circles, clearly showing the sinusoidal oscillatory waveform typical of the acoustic response. Fitting with a sum of co-sinusoidal terms results in two dominating phonon modes. Their frequencies ($\nu_S\approx\qty{43}{\THz}$ and $\nu_{LA}\approx\qty{123.0}{\THz}$) are compatible with the expected frequencies of the SAW and longitudinal acoustic (LA) phonon, respectively. A more detailed discussion of the thermoelastic properties of the sample and the fitting procedure is given in the Supplemental Material \cite{Suppl}.

We compare this signal with the one obtained with OP pump beams, plotted as red triangles in Figure \ref{Fig2}(b). The two dynamics strongly differ up to \qty{10}{\ps}, since the latter signal exhibits a clear non-sinusoidal behavior with a \qty{2}{\ps} rise and a sub-\qty{10}{\ps} decay. Qualitatively, these dynamics strongly resemble the magnetic TG signal observed in Ref. \cite{Ksenzov2021}. This is evidenced by the comparison with the gray dotted trace in panel (a) and (c). Following this decay, the signal exhibits oscillatory dynamics, suggesting the presence of a residual density modulation as discussed below.

The early-time dynamics, shown in Fig. \ref{Fig2}(c), confirm the different functional dependence of the signals, normalized to their maximum. As discussed above, the disappearance of the sinusoidal signal from the surface displacement $u_z$, and the persistence of a signal associated with a refractive index variation of magnetic nature $\Delta n_M(x, \Delta t)$, is consistent with the working hypothesis of the generation of a nanoscale transient polarization grating. 

\begin{figure}
 \includegraphics{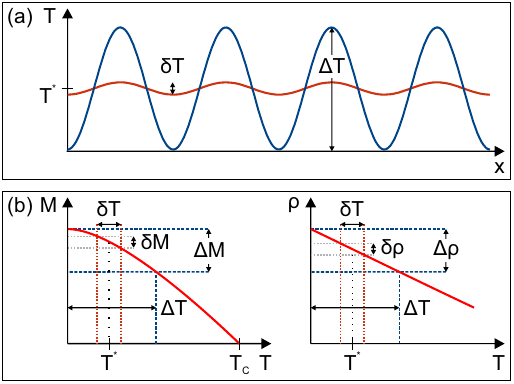}%
 \caption{\label{Fig3}(a) Cartoon of the temperature gradient $\Delta T$ induced by intensity TG compared with the residual temperature modulation $\delta T$ around $T^*$ associated with dichroic absorption in polarization TG. (b) Schematic representation of the changes in magnetization $\Delta M$ and density $\Delta \rho$ driven by the temperature modulation $\Delta T$ and the corresponding $\delta M$ and $\delta \rho$ due $\delta T$}
 \end{figure}

The dashed line in panel (b) is the fit to the backward-diffracted signal from the intensity grating scaled by $\alpha = 0.019$ (See Supplemental Material \cite{Suppl}). This residual intensity contrast can have a two-fold origin: i) a non-perfect orthogonal polarization of the two FEL pulses (See Appendix) and ii) a remaining temperature grating due to dichroic absorption $\Delta\beta$, i.e. the dichroic contribution to the imaginary part of the EUV refractive index: $n^{\pm}=(1- \delta\pm\Delta\delta)+i(\beta\pm\Delta\beta)$. 
As an order of magnitude estimate of the second mechanism, we write the amplitude of the fully modulated thermal grating resulting from intensity TG excitation as $\Delta T \propto 4I_{0,PP}/L_{abs}\sigma_{ex}^2$, where $\sigma_{ex}$ and $L_{abs}=\lambda_{ex}/4\pi\beta$ are the FEL spot size and the average absorption lengths, respectively. In the case of the polarization grating, the residual temperature grating has an amplitude $\delta T \propto 2I_{0,OP}/\sigma_{ex}^2 (1/L_{abs}^- - 1/L_{abs}^+)$ around the average temperature $ T^*\propto 2I_{0,OP}/L_{abs}\sigma_{ex}^2$. Here $L_{abs}^{\pm}=\lambda_{ex}/4\pi(\beta\pm\Delta\beta)$ account for the different absorption lengths for left and right circular polarization. The two temperature modulations are schematically depicted in Figure \ref{Fig3}(a). Since a) the spot size and overlap conditions are the same for intensity and polarization gratings, b) the amplitude of the $u_z$ grating is linearly proportional to the temperature modulation and c) the TG signal is proportional to $u_z^2(x,\Delta t)$, the ratio between the acoustic (sinusoidal) signal in intensity and polarization TGs can be written as $(\Delta T/\delta T)^2=(I_{0,PP}/I_{0,OP})^2(\beta/\Delta\beta)^2\approx 150$. This is on the same order of magnitude as the residual thermoelastic contribution extracted from the data analysis \cite{Suppl}.
 
In order to further elaborate on the nature of the helicity-dependent excitation mechanism, we first recall the interpretation of the nanoscale magnetic TG data from Ref. \cite{Ksenzov2021}. On an initial, ultrafast time scale, the intensity TG excitation modulates the electronic temperature and thereby also the spin temperature, leading to ultrafast demagnetization of the photoexcited stripes. This results in a modulation of the magnetic circular dichroism $\delta n^\pm = \pm(\Delta\delta +i\Delta\beta)\frac{\Delta M}{2}\cos{qx}$ and consequent observation of the diffracted signal. The signal grows on a timescale in agreement with the literature on ultrafast demagnetization of \ce{Co} \cite{Bergeard2014,Lopez-Flores2013} before it decays exponentially within \qty{10}{\ps}. Due to the dependence on $\Lambda_{TG}$, the decay was attributed to thermal diffusion washing away the magnetic contrast. 

Similar temporal regimes are distinguished in the case of the polarization TG and are evidenced respectively by the different shaded areas in Fig. \ref{Fig2}(c). The longer time scale is associated, also in this case, to the decay of the magnetic contrast once the magnetic grating is established.  

More intriguing are the initial and intermediate regimes in Fig. \ref{Fig2}(c), where the chiral response and subsequent growth of the magnetic contrast occur.
Several concurrent processes can lead to an helicity-dependent change of magnetization within this initial timescale. Any process inducing a temperature contrast $\delta T$ after a sub-ps thermalization, such as dichroic absorption, can be ruled out. Indeed, as sketched in Fig. \ref{Fig3}(b), $\delta T$ not only leads to a demagnetization $\delta M$ but also to a change in density $\delta \rho$, which results in the excitation of the SAW. For small temperature variations, both $\delta M$ and $\delta \rho$ can be approximated to be linear with temperature. Hence, one could expect to observe a time-dependent signal resembling the intensity TG response, scaled to the much smaller induced $\delta T$.

However, we do not observe such behavior. Therefore, the mechanism behind the observed ps-dynamics in the polarization TG must be related to the magnetization dynamics, but not to the electronic and lattice temperature.
This could be the case, for example, of MILA, a purely optical effect that is due to electron-photon scattering, and which induces, via spin-orbit coupling, transitions between electronic states of opposite spins \cite{Scheid2019,Scheid2021}. Alternatively, recent works attributed to the IFE helicity-dependent magnetization dynamics across the Fe M-edge even for photon energies where $\delta \beta \approx 0$ \cite{Hennecke2023}. Moreover, in the case of the polarization TG, we can expect a cooperation between the thermally driven changes in the magnetization induced by the uniform heating and the magnetic torque induced by the circularly polarized periodic pattern \cite{Stanciu2007}.
Although a complete understanding of the underlying excitation mechanism goes beyond the scope of the current demonstrative experiment, these results already reveal how fluence, wavelength or pulse-length dependent studies comparing intensity and polarization TGs on the same sample and in the same experimental conditions could provide relevant insights on the intricate physics of light-induced nanoscale magnetic dynamics.

In conclusion, we applied a tailored FEL setting and the unique TG instrument available at FERMI to demonstrate the generation of an ultrafast polarization grating with a \qty{43.6}{\nm} period, by selectively switching off the thermoelastic response of a magnetic system. We showed how this method is ideal to investigate ultrafast light-induced magnetization changes at the nanoscale, where the comparison of intensity and polarization TG allows to differentiate between absorption- and helicity-dependent processes. Indeed, the use of EUV pulses does not only allow to access the nanoscale but also provides the unique playground to tune the relative weight of absorption, dispersion and circular dichroism around resonances. Our approach is not limited to magnetic systems but can be extended to all systems that have chiral response, both in soft and hard condensed matter.
Examples include experiments targeting element-specific dichroic absorption in chiral molecules \cite{Mincigrucci2023,Zhang2017} and molecular crystals \cite{Peacock2001}. In other examples, the polarization grating can be used to access processes that would be otherwise hidden, as it is the case of electron drift mobility or spin Coulomb drag in semiconductor quantum wells \cite{Cameron1996, Weber2005} or exciton spin relaxation in quantum dots \cite{Scholes2006}. Finally, we anticipate that polarization TG experiments could become a unique tool to investigate chiral opto-magnetic properties and the valley degree of freedom of emerging topological systems such as Weyl semimetals \cite{Yan2017}, metal dichalchogenides \cite{Schaibley2016} or twisted bilayer graphene \cite{Liu2020}.\\
\begin{acknowledgments}
K.Y. and C. v. K. S. acknowledge financial support by the Deutsche Forschungsgemeinschaft (DFG, German Research Foundation) – Project-ID 328545488 – TRR 227, project A02.
E. P. acknowledges funding from the European Union’s Horizon 2020 research and innovation programme under the Marie Skłodowska-Curie grant agreement No 860553. B.W. and S.B. acknowledge support from the European Research Council, Starting Grant 715452 “MAGNETIC-SPEED-LIMIT”.
\end{acknowledgments}

\appendix
\section{Accelerator setup}

\begin{figure}[]
\includegraphics{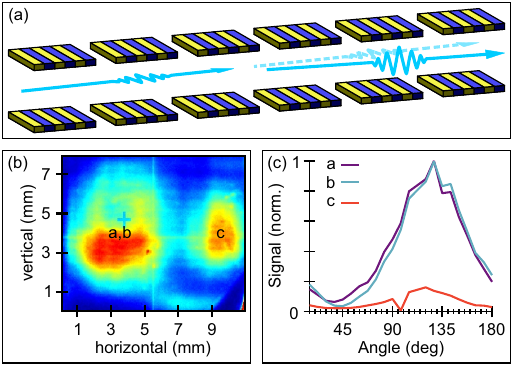}%
\caption{\label{Fig_machine} (a): Configuration of the FERMI FEL radiator, allowing the generation of two pulses  with tilted trajectory, same wavelength, balanced intensity and variable (parallel or orthogonal) linear polarization.
Fig. (b): CCD image of the two pulses, recorded about \qty{70}{\m} downstream of the last radiator. Right (a,b) linear horizontal polarization; left (c) vertical polarization. The separation between the centers of the two spots is about 5 mm.
(c): Polarization measured at the beamline for the LV beam with the LH set of undulators tuned (a, violet) and detuned (b, light blue). Polarization of the LH beam with the LH set of undulators detuned (c, red) showing the residual LV emission along that trajectory. }
\end{figure}

The FERMI free electron laser was tuned in an \textit{ad-hoc} tailored setup aiming at delivering orthogonal polarized pulses displaced in the horizontal plane.  
Figure \ref{Fig_machine}(a) schematically depicts the machine geometry. The undulator chain of FERMI FEL 1 \cite{Allaria2012} was divided in two sections of 3 undulators each that were set to a different, orthogonal, linear polarization \cite{Ferrari2019}. Additionally, the trajectory of the electron bunch is tilted to separate the pulses horizontally \cite{MacArthur2018}. Both pulses were emitted off-axis with respect to the undulator line with the idea of partially compensating the reduction of FEL gain and balancing the relative pulse energy of the two beams. 

Figure \ref{Fig_machine}(b) shows a CCD image of the two spots acquired \qty{70}{m} after the end of the undulator chain, with the linear vertical (LV) polarized beam on the left and the linear horizontal (LH) one on the right. 
The degree of polarization of the two pulses has been measured at the beamline using a polarimeter consisting of a mirror operating at the Brewster angle and a photodiode, both rotating azimuthally around the beam axis. The geometry was such that the maximum reflectivity for LH polarization corresponds to an angle of \ang{45} and for LV to \ang{135}. It is important to remark that these measurements have been performed posterior to the experiment, albeit with a similar machine configuration. The results are shown in Figure \ref{Fig_machine}(c). The violet curve (a) has been measured with the polarimeter placed along the LV trajectory and \textit{both} undulator sections tuned, i.e. both LH and LV beams were emitted by the FEL. 
The light blue curve (b) was measured with the polarimeter in the same position but only the first section of the undulators tuned, i.e. only the LV beam was emitted. Finally, to quantify the background LV light propagating collinear to the LH beam the polarimeter was centered on the second beam \textit{but} the LH emission was kept detuned. All traces have been normalized to the same factor (\num{7.13e3}), corresponding to the counts at the maximum of trace (a). The left spot in fig. 2(b) showed an almost pure vertical polarization, with the minimum of trace (b) being centered exactly on \ang{45} and the one of trace (a) slighly shifted towards lower angles. The difference of intensity between the two traces is at most \qty{6}{\percent}. The LV contamination along the LH trajectory instead was found to be \qty{15}{\percent}. 

\bibliography{biblio.bib}


\include{supplementary.tex}

\end{document}


\title{Supplemental Material: Nanoscale transient polarization gratings}

\maketitle
\onecolumngrid

\subsection{Sample properties}

The magnetic layer stack consisting of  \ce{Ta}(\qty{1.5}{nm})\slash \ce{Pt}(\qty{1.5}{nm})\slash \ce{Gd_{0.22}Co_{0.78}}(\qty{15}{nm})\slash \ce{Pt}(\qty{3}{nm}) has been grown on a selected float glass substrate using dc- and rf-magnetron sputtering at room temperature (base pressure of $\qty{3e-8}{\milli\bar}$). Ar is used as a sputter gas at a pressure of $\qty{3.4e-3}{\milli\bar}$. Typical deposition rates range between \qtylist{0.1;0.4}{\ampere\per\second}.

The CoGd layer exhibits an out of plane anisotropy with a coercive field of approximately \qty{70}{\milli\tesla}. The compensation temperature is above room temperature, which is why the magnetization for the presented measurements is dominated by Gd. Similar to other work we expect a Curie temperature larger than \qty{600}{\K} \cite{Ceballos2021} and around \qty{900}{\K}\cite{Shirakawa1985}.
The \qty{3}{nm} \ce{Ta} capping layer transmits \qty{89}{\percent} of the incident light at \qty{20.8}{\nm} and we calculate a penetration depth of \qty{16.5}{\nm} in the ferrimagnetic layer.

\subsection{Data analysis and thermoelastic properties}

Standard FEL configurations (single beam at the FEL output and circular polarization of the pump pulse) with single emission wavelength at \qty{20.8}{\nm} and double emission wavelength at \qtylist{41.6;20.8}{\nm} were used for reference EUV TG measurements. In the former case $\Lambda_{TG}=\qty{43.6}{\nm}$ ($\lambda_{ex}=\qty{20.8}{\nm}$) while in the latter $\Lambda_{TG}=87.2$ nm ($\lambda_{ex}=\qty{41.6}{\nm}$), in both cases $\lambda_{pr}=\qty{20.8}{\nm}$. Such reference measurements were carried out during the same experiment, on exactly the same sample and experimental geometry. In this case the three pump beams were split via wavefront division beam-splitting in the standard beamline setup \cite{Mincigrucci2018}.

The data obtained with an intensity grating excitation (Parallelly-polarized, PP, beams and standard FEL configurations) has been fitted with Eq. \ref{Seq1}
\begin{equation}
\label{Seq1}
I(\Delta t)  =\vert S_{ele} + S_{TE} \vert^2 \approx A^2_{ele}\,e^{-(\frac{\Delta t}{\sigma})^2} + \Theta(\Delta t)\Bigl|A_T\,e^{-\frac{\Delta t}{\tau_T}} - \sum_i A_{i}\cos(2\pi\,\nu_{i}\,(\Delta t))\,e^{-\frac{\Delta t}{\tau_{i}}}\Bigr|^2,
\end{equation}
where $S_{ele}$ and $S_{TE}$ are respectively the electronic and thermoelastic signal. 
$\Delta t$ is the time delay, $A_{ele}$ is the amplitude of the electronic signal, $\sigma$ is the experimental time resolution, $\Theta(\Delta t)$ is the Heaviside function, and $A_T$ and $\tau_T$ the amplitude and decay time of the thermal component. Finally, $A_i$, $\nu_i$ and $\tau_i$ are the amplitude, frequency and decay time of the i-th phonon mode.   

The rationale behind Eq. \ref{Seq1} is as follows:
\begin{description}
\item[$S_{ele}$] Absorption of an EUV photon is mainly followed by the creation of excited electrons with comparable energy (10s-100s of \unit{\eV}) and core-hole states. These high energy excitations relax into lower-energy electrons and shallow holes mainly via scattering and Auger recombination, two mechanisms typically faster than the employed FEL pulse duration \cite{Bencivenga2023}. In the case of an intensity TG excitation these electronic populations have the same spatial periodicity as the TG and lead to the diffraction of the EUV probe as long as their presence induces a tangible value of $\Delta n$ at $\lambda_{pr}$. This is the case of the present work, where $\lambda_{pr}$ is matched to a core resonance. The time-dependence of the EUV TG signal due to \unit{\fs}-electron dynamics can be rationalized as the convolution of a Gaussian function and an ultrafast exponential decay starting at time zero. However, since the focus of the work is the \unit{\ps} dynamics, the time zero region was coarsely sampled (or not sampled at all) and $S_{ele}$ can be reasonably approximated by a Gaussian function whose line-width $\sigma \approx \qty{95}{fs}$ FWHM represents the experimental time resolution, limited by the large spot size and crossing angle of the FEL pulses that generate the TG \cite{Foglia2023}.  
\item[$S_{TE}$]Then, on a hundreds of \unit{\fs} timescale, the population of low energy electrons thermalizes with the lattice via electron-phonon coupling. Since the electron mean free path on such a short timescale is much smaller than the TG periodicity, this results in a thermal grating of equal spacing and high visibility. This thermal grating is converted into a density grating via thermal expansion and magnetization grating via temperature-dependent magnetization (see Fig. 3 in the main text). The density grating is responsible for the thermoelastic signal in the ps regime, which, in the backward diffraction geometry employed here, is essentially proportional to $u_z^2(\Delta t)$ \cite{Maznev2021,Foglia2023,Bencivenga2023}. The resulting dynamics can be decomposed as the sum of a slow relaxation of the thermal grating, which is washed out mainly via thermal diffusion in plane with a rate $\tau_T^{-1}= 4\pi^2 D_T/\Lambda_{TG}^2$, where $D_T$ is the thermal diffusivity, modulated by the propagation of acoustic phonons, described by the the sum of damped cosinusoidal terms with a decay rate $\tau_i$ and frequency $\nu_i = c_i / \Lambda_{TG}$, where $c_i$ is the phonon propagation velocity.
\end{description}
Both electronic and thermoelastic responses give rise to signals having the same polarization as the probe beam (linear and othogonal to the scattering plane) that should be added in amplitude and then squared at the intensity detector. However, the cross-product terms in Eq. \ref{Seq1} can be safely neglected in light of the very different characteristic timescales of $S_{ele}$ and $S_{TE}$.

\begin{figure}
\centering\includegraphics[width=12 cm]{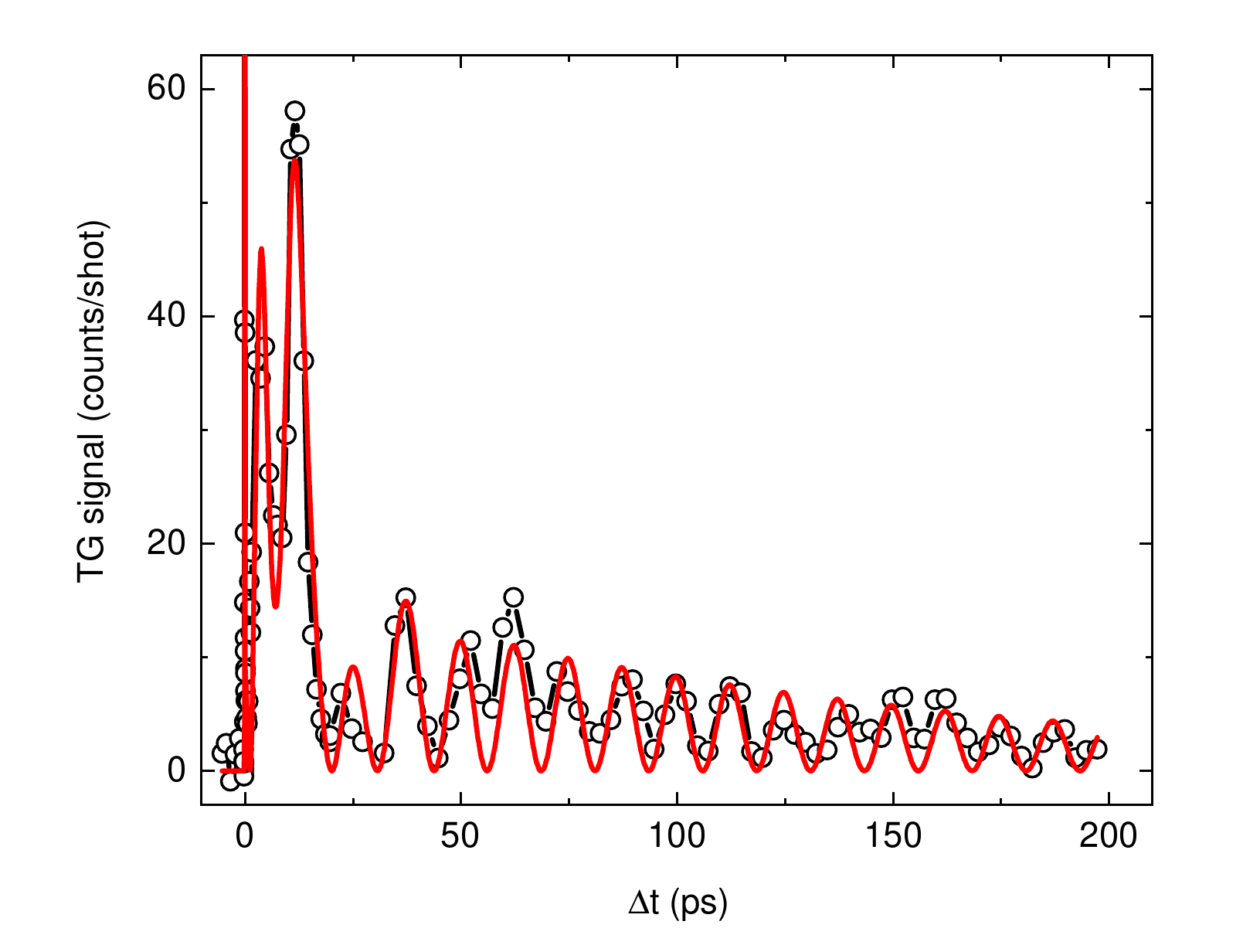}%
\caption{\label{Fig_fitPP} The red line is the best fit result of the intensity TG data collected with PP beams at $\Lambda_{TG}=43.6$ nm, represented by the black circles connected by black lines.}
\end{figure}

The result of the best fit, based on a Levenberg-Marquardt algorithm, of the waveform resulting from intensity TG excitation at $\Lambda_{TG}=\qty{43.6}{\nm}$ is displayed in Figure \ref{Fig_fitPP}, obtained considering two phonon modes in Eq. \ref{Seq1}. Note that we report here the full acquired waveform (up to $\Delta t = \qty{200}{\ps}$), while in the main text we show only the first \qty{25}{\ps} for sake of comparison with the polarization TG data.

As expected, the resulting frequency of $\nu_{SAW}=\qty{40.1\pm0.2}{\giga\Hz}$ is compatible with a weakly damped SAW mode polarized orthogonally to the sample surface. This mode is the one dominating the signal, in particular in the $\Delta t$ range of \qtyrange[range-phrase = --,range-units = single]{50}{150}{\ps}, where the waveform is very smooth and clearly features a single vibrational mode.
In the first \qty{50}{\ps} range the signal waveform is less regular and exhibits a beating with a second, highly damped, mode of frequency $\nu_{LA}=\qty{125\pm2}{\giga\Hz}$, attributed to a longitudinal acoustic (LA) phonon. While in bulk solids a strong damping of LA modes in this wavevector range is not expected, the situation can be markedly different in thin layered samples where surfaces and interfaces can be regarded as strong defects of the otherwise continuous structure. The substantial temperature rise due to the FEL pump (estimated to be about \qty{400}{\K}) can also increase the anhamrmonicity and consequent damping. Moreover, while the in-plane wavevector is well defined by the TG, the perpendicular component (z-axis) of the wavevector is not, due to the strong EUV absorption. This can launch coherent phonon wavepackets with broad spectrum in $q_z$ (and hence $\nu$), which quickly dephase due to dispersion, resulting into an apparent damping.
We stress that the elucidation of such a strong damping goes beyond the scope of this work. Here we are interested in an empirical but quantitative modeling of the thermoelastic response observed under PP excitation conditions, in order to quantify its disappearance when we move to the polarization grating.

\begin{figure}
\centering\includegraphics[width=12 cm]{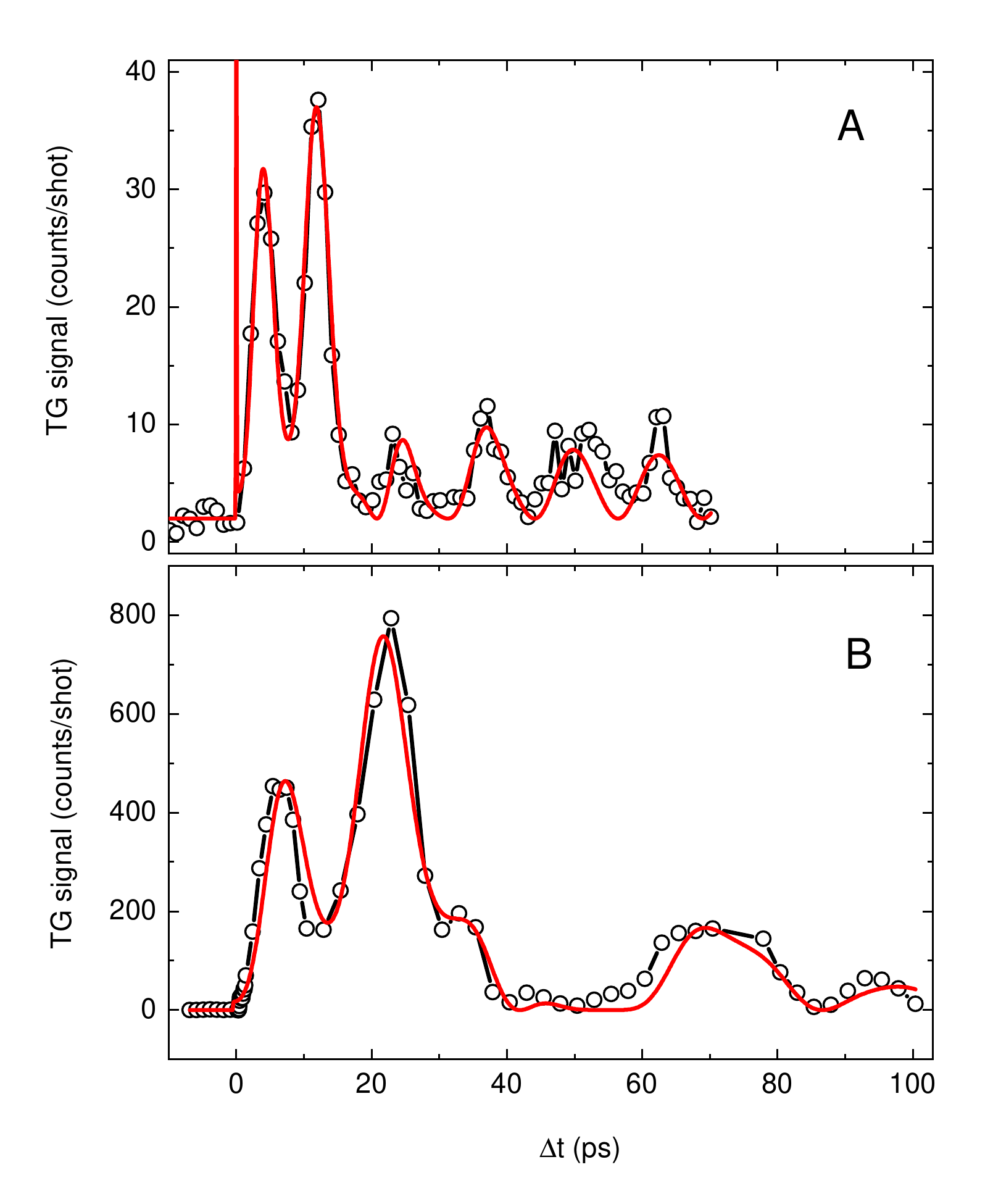}%
\caption{\label{Fig_fitRef} Red lines are best fit results of reference EUV TG data (intensity grating), represented by black circles connected by black lines, collected at $\Lambda_{TG}=43.6$ nm (panel A) and $\Lambda_{TG}=87.2$ nm (panel B).}
\end{figure}

The same approach was used to describe the EUV TG measurements carried out in standard conditions (single beam with circular polarization at the FEL output) at $\Lambda_{TG}=$\qtylist{43.6;87.2}{\nm}. The results are shown respectively in Figure \ref{Fig_fitRef} A and B; also in these cases two phonon modes are considered. The best fit results are summarized in Table \ref{tableFit}. 

\begin{table}
	\begin{tabular}[t]{|c|c|c|c|c|c|c|}
		\hline 
		Dataset & $\Lambda_{TG}$ [\unit{\nm}] & $\tau_T$ [\unit{\ps}] & $\nu_{SAW}$ [\unit{\giga\Hz}] & $\tau_{SAW}$ [\unit{\ps}] & $\nu_{LA}$ [\unit{\giga\Hz}] & $\tau_{LA}$ [\unit{\ps}] \\
		PP & 43.6 & 8.3(0.6) & 40.1(0.2) & 270(70) & 125(2) & 5(0.5) \\
		\hline
		Std & 43.6 & 7.5(0.3) & 40.0(0.5) & 270(fix) & 124(2) & 13(2) \\
		\hline
		Std & 87.2 & 23.9(1.0) & 20.5(0.5) & 1000(fix) & 70.0(1.4) & 30(7) \\
		\hline 
		
	\end{tabular}
 	\caption{Best fit values for the parameters in the $I_{TE}$ term of Equation \ref{Seq1}; errors in the last significant digits are in brackets. "Std" stands for standard conditions. Because of the limited range in $\Delta t$, the value of $\tau_{SAW}$ for the standard measurements at $\Lambda_{TG}=\qty{43.6}{\nm}$ was fixed to the one found in the fitting of the data acquired with PP beams. The value of $\tau_{SAW}$ for the standard measurements at $\Lambda_{TG}=\qty{87.2}{\nm}$ was fixed to \qty{1}{\ns} since this mode did not show an appreciable decay in the probed $\Delta t$ range (\qtyrange[range-phrase = --,range-units = single]{0}{100}{\ps}). \label{tableFit}}
\end{table}%

\begin{figure}
\centering\includegraphics[width=12 cm]{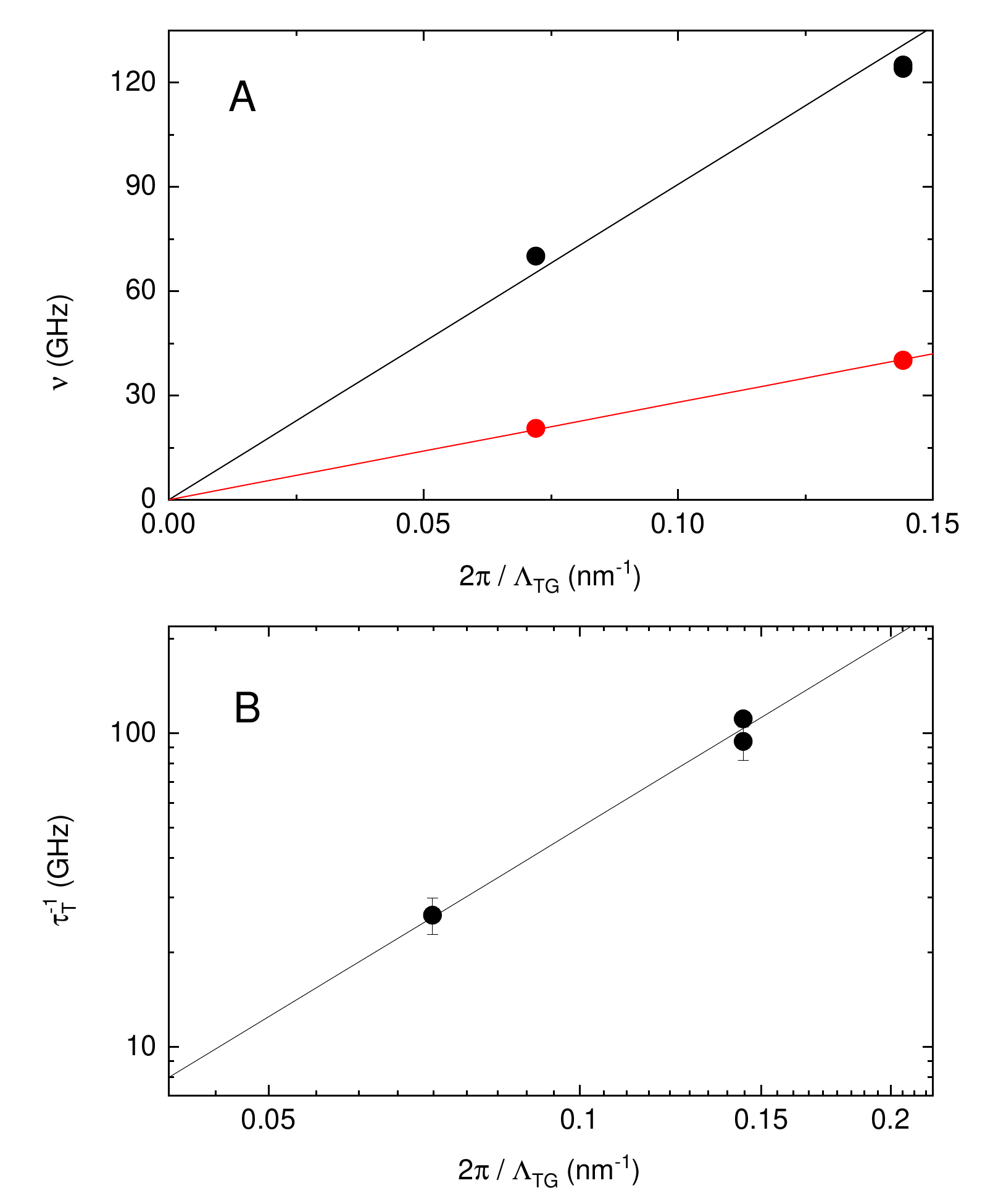}%
\caption{\label{FitPara} Panel A: values of $\nu_{SAW}$ (red dots) and $\nu_{LA}$ (black dots) as a function of $q=2\pi/\Lambda_{TG}$, as extracted from the best fit procedure. Straight lines through the data represent a velocity of $c_{SAW}= \qty{1760}{\m\per\s}$ and $c_{LA}= \qty{5700}{\m\per\s}$, respectively. Panel B: values of $\tau_T^{-1}$ (black dots) as a function of $q$ in a semi-logarithmic plot, the black line correspond to $D_T=\qty{5}{\square\nm\per\ps}$.}
\end{figure}

The linear phonon dispersion as a function of $q=2\pi/\Lambda_{TG}$ is plotted in Fig. \ref{FitPara} A. The SAW (red dots and line) is associated with a velocity $c_{SAW}= \qty{1760}{\m\per\s}$, matching values expected for SAW in samples similar to the used one. The propagation velocity $c_{LA}= \qty{5700}{\m\per\s}$ extracted for the second mode (black dots and line) confirms its LA nature. 
The obtained values for the decay rate of the thermal grating as a function of $q$ are displayed in Figure \ref{FitPara} B and are compatible with a value of $D_T=\qty{5}{\square\nm\per\ps}$. We are not aware of reference data for $D_T$ in \ce{CoGd} alloys, while the calorimetric data, corresponding to macroscopic ($\Lambda_{TG}\rightarrow\infty$) length-scales, for \ce{Co} and \ce{Gd} are $D_T=\qtylist{22.7;4.9}{\square\nm\per\ps}$, respectively. 

\begin{figure}
\centering\includegraphics[width=12 cm]{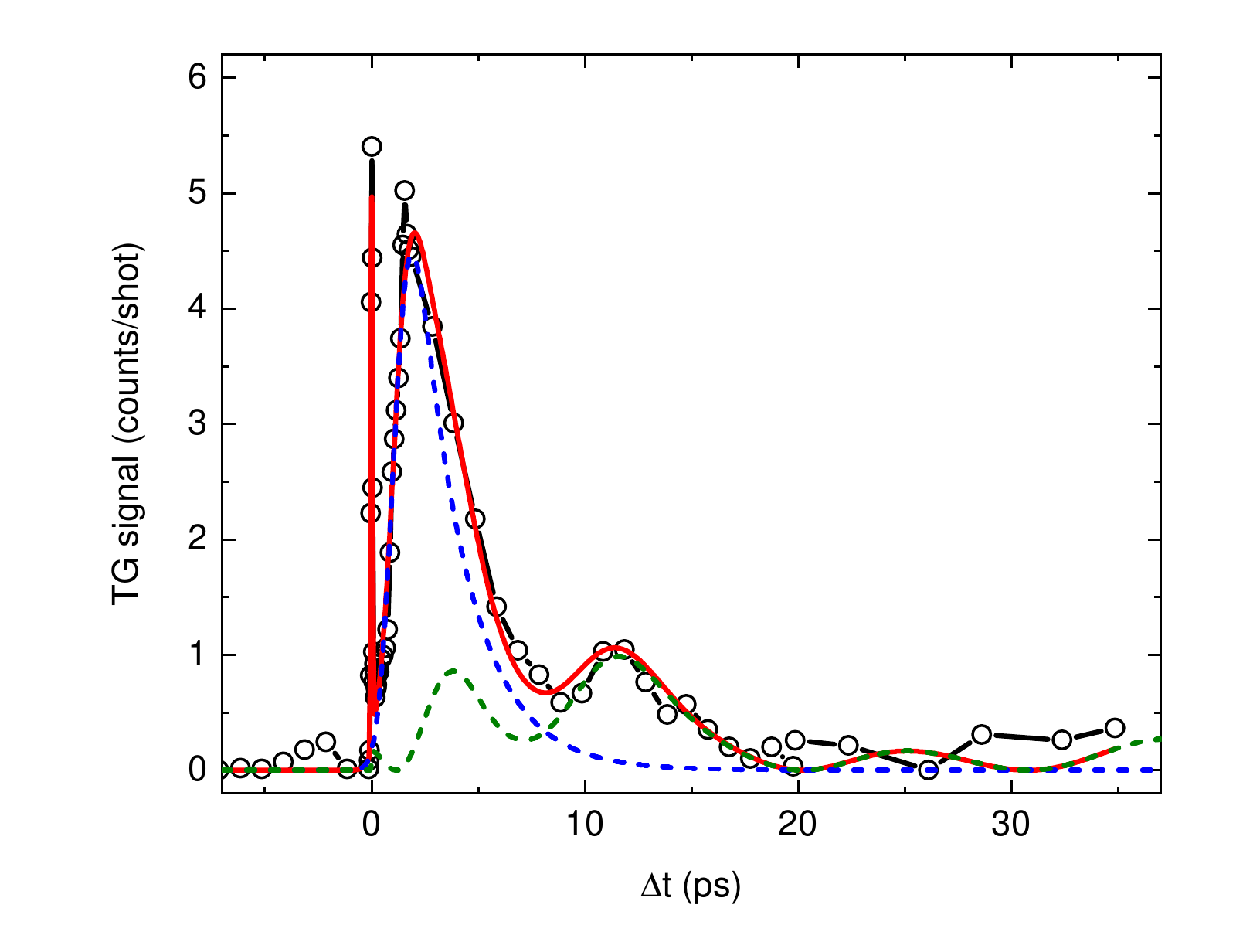}%
\caption{\label{CrossPol} The red line is the best fit result of the polarization grating data collected at $\Lambda_{TG}=\qty{43.6}{nm}$, represented by the black circles connected by black lines. Blue and green dashed lines are, respectively, the $\alpha\vert S_{TE}^{PP}\vert^2$ and $\vert S_{mag}\vert^2$ terms in Eq. \ref{Seq2}.}
\end{figure}

Eq. \ref{Seq1} cannot be used to describe the strongly non-sinusoidal shape of the polarization grating data, hereafter referred to with the notation OP. However, as explained in the main text, the oscillation at $\Delta t\approx \qty{11}{\ps}$ in that waveform suggests a small residual contribution of the thermoelastic response, most likely arising from a weak intesity grating due to dichroic absorption. Moreover, since the magnetic response rotates the polarization of the probe beam \cite{Ksenzov2021}, the signal $S_{mag}$ arising from a magnetization grating is polarized orthogonally with respect to $S_{ele}$ and $S_{TE}$ and Eq. \ref{Seq1} becomes:

\begin{equation}
\label{Seq2}
I_{OP}(\Delta t) = \vert S_{ele} + S_{TE} \vert^2 + \vert S_{mag} \vert^2.
\end{equation}

Applying the same reasoning as above for the cross-product terms of $S_{ele}$ and $S_{TE}$, the polarization grating data are then fitted with the following function:

\begin{equation}
\label{Seq3}
I_{OP}(\Delta t) = \vert S_{ele}\vert^2 +\alpha\vert S_{TE}^{PP} \vert^2 + \vert S_{mag} \vert^2.
\end{equation}
Here $\alpha$ is a scaling factor, $S_{TE}^{PP}$ indicates the $S_{TE}$ function with all fitting parameters (including the amplitudes $A_T$, $A_{SAW}$ and $A_{LA}$) fixed to the values found in the the analysis of the intensity grating waveform with PP beams, and $S_{mag}$ is an exponential modified Gaussian function
\begin{equation*}
S_{mag} = \frac{A_m}{2\tau_m} e^{(\frac{t_m}{\tau_m}+ \frac{1}{2}(\frac{\sigma_m}{\tau_m})^2 - \frac{\Delta t}{\tau_m})} \erf\left(\frac{t_m + \sigma_m^2/\tau_m - \Delta t}{\sqrt(2) \sigma_m}\right)
\end{equation*}
with variance $\sigma_m$, mean $t_m$ and decay time $\tau_m$. Differently from $S_{TE}$, the choice of the functional form for the $S_{mag}$ term is empirical and based on the simplicity of the function and easiness in determining the shape parameters: the position in time of the maximum signal, the rise and decay times.
The best fit results of the polarization grating data, as obtained by using Eq. \ref{Seq3} are shown in figure \ref{CrossPol}. The best fit parameters are $\alpha=\num{0.019\pm 0.003}$, $\sigma_m\qty{0.71 \pm 0.05}{ps}$, $t_m=\qty{0.71 \pm 0.04}{ps}$ and $\tau_m=\qty{4.5 \pm 0.5}{\ps}$.

From the value of $\alpha$ and after normalization by the total intensity ($I_0$) of the FEL output, i.e. $(I_0^{OP}/I_0^{PP})^3\approx 1.8$ (in the employed conditions the EUV TG signal is proportional to $I_0^3$), one can estimate a residual amplitude of $\approx$\qty{1}{\percent} for the remaining thermoelastic signal due to the residual intensity grating, compatible with  the amplitude evaluated for the residual intensity grating due to dichroic absorption and also on the same level as the signal eventually occurring from a non perfect orthogonality in the polarization of the two FEL pulses, as evaluated in a separate experimental run in a similar FEL configuration (see appendix in the main text).

\bibliography{biblio.bib}